\documentclass[prb,twocolumn,showpacs]{revtex4}
\usepackage{graphicx,epsf}

\newcommand{\bq}{{\bf q}}
\newcommand{\bk}{{\bf k}}

\newcommand{\br}{{\bf r}}
\newcommand{\ba}{{\bf a}}
\newcommand{\be}{{\bf e}}
\newcommand{\bR}{{\bf R}}
\newcommand{\bA}{{\bf A}}
\newcommand{\bP}{{\bf P}}

\newcommand{\etab}{\mbox{\boldmath $\eta $}}
\newcommand{\Pib}{\mbox{\boldmath $\Pi $}}
\newcommand{\rhobar}{\bar{\rho}}
\newcommand{\chibar}{\bar{\chi}}
\newcommand{\rhobarbar}{\bar{\bar{\rho}}}

\newcommand{\Hhat}{\hat{H}}

\newcommand{\nablab}{\mbox{\boldmath $\nabla $}}

\newcommand{\beq}{\begin{equation}}
\newcommand{\eeq}{\end{equation}}
\newcommand{\beqn}{\begin{eqnarray}}
\newcommand{\eeqn}{\end{eqnarray}}
\newcommand{\nn}{\nonumber}

\begin{document}

\def\tende#1{\,\vtop{\ialign{##\crcr\rightarrowfill\crcr
\noalign{\kern-1pt\nointerlineskip}
\hskip3.pt${\scriptstyle #1}$\hskip3.pt\crcr}}\,}

\title{Spin-excitations of the quantum Hall ferromagnet of composite fermions}

\author{R. L. Doretto,$^{1,2}$ M. O. Goerbig,$^{2,3}$ P. Lederer,$^4$   
        A. O. Caldeira,$^1$ and C. Morais Smith$^{2,5}$}

\affiliation{$^1$Departamento de F\'{\i}sica da Mat\'eria Condensada,
                 Instituto de F\'{\i}sica ``Gleb Wataghin'', 
                 Universidade Estadual de Campinas, CEP 13083-970, 
                 Campinas-SP, Brazil\\
$^2$D\'epartement de Physique, Universit\'e de Fribourg, P\'erolles,  
    CH-1700 Fribourg, Switzerland\\
$^3$Laboratoire de Physique Th\'eorique et de Hautes \'Energies, CNRS UMR 7589,
    Universit\'es Paris 6 et 7, 4, Place Jussieu, F-75252 Paris cedex 05, 
    France\\
$^4$Laboratoire de Physique des Solides, B\^at.\,510 (associ\'e au CNRS),
    Universit\'e Paris-Sud, F-91405 Orsay cedex, France\\
$^5$Institute for Theoretical Physics, University of Utrecht, Leuvenlaan 4,
    3584 CE Utrecht, The Netherlands}

\begin{abstract} 
The spin-excitations of a fractional quantum Hall system are evaluated
within a bosonization approach. In a first step, we generalize Murthy and 
Shankar's Hamiltonian theory of the fractional quantum Hall effect to the 
case of composite fermions with an extra discrete degree of freedom. Here, we
mainly investigate the spin degrees of freedom, but the proposed formalism may
be useful also in the study of bilayer quantum-Hall systems, where the layer 
index may formally be treated as an isospin. In a second step, we 
apply a bosonization scheme, recently developed for the study of the 
two-dimensional electron gas, to the interacting composite-fermion
Hamiltonian. The dispersion of the bosons, which represent 
quasiparticle-quasihole excitations, is analytically evaluated for 
fractional quantum Hall systems at $\nu = 1/3$ and $\nu = 1/5$. 
The finite width of the two-dimensional electron gas is also taken into
account explicitly. Furthermore, we consider the interacting bosonic model and
calculate the lowest-energy state for two bosons. In addition to a continuum 
describing scattering states, we find a bound-state of two bosons. This state 
is interpreted as a pair excitation, which consists of a  
skyrmion of composite fermions and an antiskyrmion of composite fermions.
The dispersion relation of the two-boson state is evaluated for $\nu = 1/3$ 
and $\nu = 1/5$. Finally, we show that our theory provides the microscopic 
basis for a phenomenological non-linear sigma-model for studying the 
skyrmion of composite fermions. 
\end{abstract}
\maketitle

\section{Introduction}

Quantum Hall physics -- the study of two-dimensional (2D) electrons in a strong
magnetic field -- has revealed a lot of unexpected phenomena during the last 25
years.\cite{perspectives} Apart from the integer and fractional quantum 
Hall effects (IQHE and FQHE, respectively), which have been identified as 
macroscopic quantum phenomena, exotic topological spin excitations are displayed
by these systems.
The physical properties of quantum Hall systems are governed by the Landau
quantization; the energy of 2D electrons with a band mass $m_B$ and charge $-e$
in a perpendicular magnetic field $B$ is quantized into equidistant energy 
levels, the so-called Landau levels (LLs), with a level separation 
$\hbar w_C = \hbar eB/m_B c$. The large LL degeneracy is characterized by the 
flux density $n_\phi=1/2\pi l_B^2$, given in terms of the magnetic length 
$l_B=\sqrt{\hbar c/eB}$, and
the LL filling is thus defined as the ratio $\nu=n_{el}/n_\phi$ of the
electronic density $n_{el}$ and $n_\phi$. Due to the Zeeman effect, each LL $n$ 
is split into two spin-branches with an energy separation $g^*\mu_B B$, where
$g^*$ is the effective Land\'e factor ($g^*=-0.44$ for GaAs), and 
$\mu_B=2\hbar e/m$ is the Bohr magneton. Because the band mass is reduced with
respect to the bare electron mass $m$ ($m_B=0.068m$ for GaAs), the spin-branch
separation is about $70$ times smaller than the LL separation. 

The IQHE may be understood in a one-particle picture, in which the
Coulomb interaction is only a small perturbation. If $\nu=N$, with integral 
$N$, the ground state
is non-degenerate and one has to provide a finite energy to promote an electron
to the next higher level. Due to the localization of additional electrons by 
residual impurities in the sample, the Hall resistance
remains at its quantized value $R_H=h/e^2N$ over 
a certain range of the magnetic
field. This plateau in the Hall resistance is accompanied by a vanishing 
longitudinal resistance, and both are the signature of the quantum Hall 
effect. If the lower
spin branch of the $n$-th LL is completely filled, one has $N=2n+1$, and 
$N=2n+2$ in the case of complete filling of both spin branches. 

The one-particle picture ceases to be valid at partial filling $\nu\neq N$,
where one is confronted with the LL degeneracy. In this limit, electrons
in a partially filled level are strongly correlated due to their mutual 
Coulomb repulsion, which constitutes the relevant energy scale. In the lowest
LL, the formation of composite fermions (CFs) leads to incompressible quantum
liquids at $\nu=p/(2sp+1)$.\cite{jain,heinonen} The FQHE, displayed by these
liquids, may be interpreted as an IQHE of CFs. The physical properties of CFs
have been extensively studied in the framework of Jain's wave-function 
approach,\cite{jain,heinonen} which is a generalization of Laughlin's trial
wave functions.\cite{laughlin} A complementary field-theoretical formalism, 
based on Chern-Simons transformations, has been proposed by Lopez and 
Fradkin.\cite{lopezfradkin} The latter has been particularly successful in
the description of the metallic state at $\nu=1/2$, which may be interpreted as
a Fermi sea of CFs.\cite{HLR} More recently, Murthy and Shankar have developed a
Hamiltonian theory of the FQHE, a second-quantized approach.\cite{MS} It
systematically accounts for the mechanism of how these excitations
are formed and what are their constituents, namely, electrons or holes
bound to an even number of vortex-like objects.  

Although the main features of the IQHE may be understood in the 
framework of a noninteracting model of spin-polarized electrons,
the study of spin-excitations 
at $\nu=1$, when only one spin-branch of the 
lowest LL ($n=0$) is completely filled, requires a more complete treatment 
of the problem. Indeed, the lowest-energy excitations involve
a spin reversal, and the spin degree of freedom must be considered
explicitly. In addition to spin-wave excitations, which are gapped due to
the Zeeman effect, one finds topological {\sl skyrmion} 
excitations;\cite{skyrme} if the
Coulomb interaction is large with respect to the Zeeman gap, it is 
energetically favorable to distribute the spin reversal over a group of 
neighboring spins.\cite{leekane,sondhi,moon} These skyrmion excitations give 
rise to unusual spin polarizations.\cite{barrett} 
More recently, the different spin excitations at $\nu=1$ have been 
investigated in a bosonization approach.\cite{doretto} 
Although the bosonization method of fermionic systems is well established for
one-dimensional electron systems,\cite{delf,voit} only few generalizations 
to higher dimensions have been proposed. \cite{castroneto,houghton} 
One particular example is a recently developed bosonization scheme 
for the 2D electron gas (2DEG)
in a magnetic field.\cite{harry,doretto} The extension presented in 
Ref.\ \onlinecite{doretto}, in particular, allows for the treatment of 
complex many-body structures, such as small skyrmion-antiskyrmion pairs
in terms of bound states of the
bosonic excitations.

Here, we generalize the Hamiltonian theory of the FQHE of Ref.\ 
\onlinecite{MS} to incorporate spin or any other discrete degree of 
freedom. Although in this paper we apply our theory to the
investigation of spin excitations, the developed formalism has 
a wider range of application and may also be useful for further studies on 
bilayer quantum Hall systems, for which the layer index 
is formally treated as an isospin.\cite{perspectives,moon}
For certain values of the magnetic field, the ground state may be viewed as a 
$\nu = 1$ state of CFs, and it is therefore natural to test the 
bosonization scheme for these cases. The method allows us to investigate
the CF excitation spectrum and to describe, for example,
skyrmion-like excitations
at filling factors such as $\nu = 1/3$. 
Indeed, optically pumped nuclear magnetic resonance 
measurements, which give information about the spin
polarization of the 2DEG, indicate the existence of such
excitation around $\nu=1/3$.\cite{khandelwal2} 

The outline of the paper is the following. The generalization of the 
Hamiltonian theory to the case of fermions with an extra discrete 
degree of freedom is presented in Sec.\ II. In Sec.\ III, we apply
the previously developed bosonization method \cite{doretto} to the obtained CF
Hamiltonian and study the dispersion relation of the resulting bosons. Bound 
states of two of these elementary excitations, which we interpret as 
small skyrmion-antiskyrmion pairs, are also considered.
Sec.\ IV is devoted to the semiclassical limit of our approach, from 
which we show that an effective Lagrangian, similar to the one suggested by
Sondhi {\it et al.},\cite{sondhi} accounts for the long-wavelength properties
of the system. A summary of the main results is finally presented in
Sec.\ V.

\section{Hamiltonian Theory of CFs with Spin}  

Our first aim is to derive the Hamiltonian theory for electrons
with spin in the lowest LL and construct the CF basis. In principle, the 
generalization of the algebraic properties of the Murthy and Shankar model 
[Ref.\ \onlinecite{MS}] to the case of discrete degrees of freedom, 
such as the spin or the isospin in
the context of bilayer quantum Hall systems, may be written down intuitively.
In this section, however, they are derived, with the help of a set of 
transformations, from the microscopic model of an interacting 2DEG, with
spin $\sigma=\uparrow$ (+1) or $\downarrow$ (-1) in a 
perpendicular magnetic field $B{\bf e}_z= \nablab\times \bA$.
The Hamiltonian of the system reads, in terms of the electron fields 
$\psi_{\sigma}(\br)$ and $\psi_{\sigma}^{\dagger}(\br)$ ($\hbar = c = 1$)
\beq
\label{eq001}
\Hhat = \Hhat_0 + \Hhat_Z + \Hhat_{I}, 
\eeq
where the orbital part is 
\beq
\label{eq001a}
\Hhat_0 = \frac{1}{2m_B}\sum_{\sigma}\int d^2r\psi_{\sigma}^{\dagger}(\br)
     \left[-i\mathbf{\nablab} +e\bA(\br)\right]^2\psi_{\sigma}(\br),
\eeq
the Zeeman term is 
\beq
\label{eq001b}
\Hhat_Z = \frac{1}{2}g^*\mu_B B\sum_{\sigma}\sigma\int
      d^2r\psi_{\sigma}^{\dagger}(\br)\psi_{\sigma}(\br),
\eeq
and the interaction Hamiltonian is 
\beq
\label{eq002}
\Hhat_{I}=\frac{1}{2}\sum_{\sigma,\sigma'}\int d^2rd^2r'\psi_{\sigma}^{\dagger}(\br)
\psi_{\sigma'}^{\dagger}(\br')V(\br-\br')\psi_{\sigma'}(\br')\psi_{\sigma}(\br),
\eeq
which couples electrons with different spin orientation.
Here, $V(\br)=e^2/\epsilon r$ is the isotropic Coulomb interaction potential, 
with $r=|\br|$, and $\epsilon$ is the dielectric constant of the host 
semiconductor.

\subsection{Microscopic Theory}

Following closely the steps of Ref.\ \onlinecite{MS}, 
we perform a Chern-Simons transformation on the electron fields, defined by
\beq
\label{eq003}
\psi_{\sigma}(\br)=e^{-2is_{\sigma}\int
  d^2r'\theta(\br-\br')\rho_{\sigma}(\br')}\psi_\sigma^{CS}(\br),
\eeq
where $\theta(\br)$ is the angle between the vector $\br$ and the $x$-axis. The
density operator reads 
$\rho_{\sigma}(\br)=\psi_{\sigma}^{\dagger}(\br)\psi_{\sigma}(\br)=
\psi_{\sigma}^{CS\dagger}(\br)\psi_{\sigma}^{CS}(\br)$. Thus, the Zeeman 
$\Hhat_Z$ and the interaction $\Hhat_{I}$ terms in Eq.\ (\ref{eq001}) remain
invariant under the above transformation. In order to 
describe fermionic fields, which is the case of interest here, $s_{\sigma}$ 
must be integer. Note that this transformation breaks the $SU(2)$ symmetry 
explicitly unless $s_\uparrow = s_\downarrow$. 
For the symmetric case, a Hamiltonian theory for bilayer systems has very 
recently been established by Stani\'c and Milovanovi\'c who investigated 
composite bosons at a total filling factor $\nu=1$.\cite{stanic} Here, we 
discuss the more general case in which the electronic density of 
$\uparrow$-electrons is not necessarily
the same as that of $\downarrow$-electrons. The transformation generates 
the two-component spinor Chern-Simons vector potential
$\ba_{\sigma}^{CS}(\br)$ with 
\beq
\label{eq004}
\nablab\times \ba_{\sigma}^{CS}(\br)=-2s_{\sigma}\phi_0\rho_{\sigma}(\br)\be_z,
\eeq
in terms of the flux quantum $\phi_0=2\pi/e$. 

In order to decouple the electronic degrees of freedom 
and the complicated fluctuations of the Chern-Simons field, Murthy and 
Shankar introduced an auxiliary vector
potential\cite{MS} $\ba_{\sigma}(\bq)$, which was inspired by the plasmon 
description proposed by Bohm and Pines for the electron gas.\cite{bohm}
If the spin degree of freedom is taken into account, this
vector potential consists of two spinor components, both of which are chosen 
transverse to satisfy the Coulomb gauge, $\nablab\cdot\ba_{\sigma}(\br)=0$. 
For its Fourier components one thus finds 
$\ba_{\sigma}(\bq)=-i\be_z\times\be_q a_{\sigma}(\bq)$, where $\be_q=\bq/|\bq|$
is the unit vector in the direction of propagation of the field. Its conjugate
field is chosen longitudinal, $\bP_{\sigma}(\bq)=i\be_qP_{\sigma}(\bq)$, with
the commutation relations
\beq
\label{eq006}
\left[a_{\sigma}(\bq),P_{\sigma'}(-\bq')\right]=
    i\delta_{\sigma,\sigma'}\delta_{\bq,\bq'}.
\eeq
The Hilbert space is enlarged by these non-physical degrees of freedom, and
one therefore has to impose the constraint $a_{\sigma}(\bq)|\phi\rangle = 0$ on
the physical states $|\phi\rangle$. The orbital part (\ref{eq001a}) of the 
Hamiltonian thus becomes
\beqn
\label{eq005}
\nn
\Hhat_0^{CS}&=&\frac{1}{2m_B}\sum_{\sigma}\int d^2r\psi_{\sigma}^{CS\dagger}(\br)\\
\nn && \\
&&\times\left[\Pib_{\sigma}+e:\ba_{\sigma}^{CS}(\br):
+e\ba_{\sigma}(\br)\right]^2\psi_{\sigma}^{CS}(\br),
\eeqn
where the mean-field part of the Chern-Simons vector potential 
$\langle \ba_{\sigma}^{CS}(\br)\rangle$ has been absorbed into a generalized
momentum $\Pib_{\sigma}=-i\nablab +e\left[\bA(\br)+
\langle \ba_{\sigma}^{CS}(\br)\rangle\right]$, and $:\ba_{\sigma}^{CS}(\br):$ 
denotes its normal-ordered fluctuations. The
conjugate field $P_{\sigma}(\bq)$ generates the translations of the vector
potential, and one may therefore choose the unitary transformation
$\psi_{\sigma}^{CS}(\br)=U_{\sigma}\psi_{\sigma}^{CP}(\br)$, in terms of 
the composite-particle field $\psi_{\sigma}^{CP}(\br)$, with
\beq 
\nn
U_{\sigma} =\exp\left(i\sum_{\bq}^QP_{\sigma}(-\bq)\frac{4\pi
  s_{\sigma}}{q}\delta\rho(\bq)\right) ,
\eeq
to make the long-range ($q < Q$) fluctuations of the
Chern-Simons vector potential vanish. The transformed Hamiltonian
(\ref{eq005}) without the short-range terms related to
$:a_{\sigma}^{CS}(\bq):$ thus reads 
\beq
\label{eq007}
\Hhat_0^{CP}=\sum_{\sigma}\int d^2r \psi_{\sigma}^{CP\dagger}
(\br)\mathcal{H}_{\sigma}(\br)\psi_{\sigma}^{CP}(\br),
\eeq
with the Hamiltonian density 
\beqn
\label{eq008}
\nn
\mathcal{H}_{\sigma}&=&\frac{\Pib_{\sigma}^2}{2m_B}+\frac{1}{2m_B}
 \left[\ba_{\sigma}(\br)+4\pi s_{\sigma}\bP_{\sigma}(\br)\right]^2\\
\nn && \\ 
  &&  +\frac{1}{m_B}\Pib_{\sigma}\cdot\left[\ba_{\sigma}(\br)+
      4\pi s_{\sigma}\bP_{\sigma}(\br)\right].
\eeqn
The second term is the Hamiltonian density of a quantum mechanical harmonic 
oscillator, as may be seen from the commutation relations (\ref{eq006}). 
In the absence of the third term of Eq. (\ref{eq008}), which couples the 
electric current density and the harmonic oscillators, one may write down
the field as a product of the ground-state wavefunction of the oscillators
$\chi(\br)$ and $\phi^*(\br)$, where $\phi^*(\br)$ is the $N$-particle 
wavefunction of charged particles in an effective magnetic field
\beq
\nn
  \nablab\times\left[\bA(\br)+\langle
  \ba_{\sigma}^{CS}(\br)\rangle\right] = B_{\sigma}^*\be_z,
\eeq
with $B_{\sigma}^*=B-2s_{\sigma}\phi_0 n_{el}^{\sigma}
=B_{\sigma}/(2s_{\sigma}p_{\sigma}+1)$, in terms of the average density 
$n_{el}^{\sigma}$ of electrons with spin $\sigma$.
In the absence of correlations between electrons with different spin,
the filling of each spin branch may thus be characterized by a renormalized
filling factor, $\nu_{\sigma}^*=n_{el}^{\sigma}2\pi l_{\sigma}^{*2}$, with
the new magnetic length $l_{\sigma}^*=1/\sqrt{eB_{\sigma}^*}$. 
The oscillator wavefunction becomes, with the help of the transformed 
constraint, $\left[qa_{\sigma}(\bq)-4\pi s_{\sigma}\rho_{\sigma}(\bq)\right]
|\phi\rangle=0$
$$\chi(\br)= \exp\left[-\sum_{\sigma,\bq}2\pi
  s_{\sigma}\delta\rho_{\sigma}(-\bq)\frac{1}{q^2}
\delta\rho_{\sigma}(\bq)\right].$$
The exponent may be rewritten in terms of the complex coordinates
$z_j=x_j+iy_j$ for spin-up and $w_\mu=x_\mu+iy_\mu$ for spin-down particles
\cite{kane} 
\beqn
\nn
\sum_{\sigma,\bq}2\pi s_{\sigma}\delta\rho_{\sigma}(-\bq)\frac{1}{q^2}
\delta\rho_{\sigma}(\bq)\simeq\\
\nn 
\sum_{i<j}2s_\uparrow\ln|z_i-z_j|-\sum_{j}\frac{|z_j|^2}{4l_B^2}c_\uparrow^2+\\
\nn
\sum_{\mu<\nu}2s_\downarrow\ln|w_{\mu}-w_{\nu}|-\sum_{j}\frac{|w_{\mu|^2}}{4l_B^2}c_\downarrow^2, 
\eeqn
where $c_{\sigma}^2=2s_{\sigma}p_{\sigma}/(2s_{\sigma}p_{\sigma}+1)$,
at integer values of $\nu_{\sigma}^*=p_{\sigma}$. The complete
wavefunction therefore becomes
\beq
\label{eq009}
\psi(z_i,w_\mu)=\prod_{i<j}\left(z_i-z_j\right)^{2s_\uparrow}
\prod_{\mu<\nu}\left(w_{\mu}-w_{\nu}\right)^{2s_\downarrow}\phi^*(z_i,w_{\mu}),
\eeq
where the Gaussian factors have been absorbed into the electronic part of the
wavefunction $\phi^*(z_i,w_\mu)$.
If there are no correlations between particles with different spin
orientation, i.e. in the absence of interactions, the electronic
part of the wavefunction may be written as a product,
$\phi^*(z_i,w_\mu)=\phi_{p_\uparrow}^\uparrow(z_i)\phi_{p_\downarrow}^\downarrow(w_\mu)$,
and Eq.\ (\ref{eq009}) thus represents a product of two Laughlin
wavefunctions\cite{laughlin} (for $p_{\sigma}=1$)
or unprojected Jain wavefunctions\cite{jain} (for $p_{\sigma}>1$) for the
two spin orientations. 
If one includes correlations between particles with
different spin orientation, one may use the ansatz 
$$
\phi^*(z_i,w_{\mu})=\phi_{p_\uparrow}^\uparrow(z_i)\phi_{p_\downarrow}^\downarrow(w_{\mu})
\prod_{i,\mu}\left(z_i-w_{\mu}\right)^m,
$$
with integral $m$. For $p_{\sigma}=1$, one recovers Halperin's
wavefunctions, \cite{halperin3} if
$$
\phi_1^{\uparrow}(z_i)=\prod_{i<j}(z_i-z_j)~
e^{-\sum_j|z_j|^2/4l_B^2},
$$
and similarly
$$
\phi_1^{\downarrow}(w_\mu)=\prod_{\mu<\nu}(w_\mu-w_\nu)~
e^{-\sum_\nu|w_\nu|^2/4l_B^2}.
$$
In this case the filling factors of the two spin branches become
$\nu_{\sigma}=(2s_{\sigma}-m)/(s_\uparrow s_\downarrow - m^2)$.
For larger values of $p_{\sigma}$, the
wavefunction (\ref{eq009}) contains components in higher LLs and is
thus no longer analytic, as required by the lowest LL condition.
\cite{jach} Eq. (\ref{eq009}) must therefore be projected into the
lowest LL, and the final trial wavefunctions become
\beqn
\label{eq010}
\nn
\psi(z_i,w_{\mu})&=&\mathcal{P}\prod_{i<j}\left(z_i-z_j\right)^{2s_\uparrow}
\prod_{\mu<\nu}\left(w_{\mu}-w_{\nu}\right)^{2s_\downarrow}\\
&&\times\prod_{i,\mu}\left(z_i-w_{\mu}\right)^m
\phi_{p_\uparrow}^\uparrow(z_i)\phi_{p_\downarrow}^\downarrow(w_{\mu}),
\eeqn
where $\mathcal{P}$ denotes the projection to the lowest LL. They may
be interpreted as a CF generalization of Halperin's
wavefunctions.\cite{halperin3}

Up to this point, the coupling term between the electronic and
oscillator degrees of freedom in the Hamiltonian density (\ref{eq008})
has been neglected. Murthy and Shankar have proposed a decoupling
procedure, which also applies to the case of electrons with spin,
with the only difference that two unitary transformations 
$\bar{U}_{\sigma}$ are required for the two spin orientations. The
reader is referred to their review (Ref.\ \onlinecite{MS}) for the details of
the decoupling transformations, and we will limit ourselves to the
presentation of the results. With the choice of equal number of
oscillators and particles for each spin orientation, the kinetic term
in the Hamiltonian vanishes, and the model in the limit of small wave
vectors becomes 
\beqn
\label{eq011}
\nn
\Hhat&=&\sum_{\sigma,\bq}\omega_CA_{\sigma}^{\dagger}(\bq)A_{\sigma}(\bq)+
\frac{1}{2}\gamma\sum_{\sigma}\sigma \rho_{\sigma}(\bq=0)\\
\nn && \\
&&+\frac{1}{2}\sum_{\bq}\sum_{\sigma,\sigma'}v_0(q)\rho_\sigma(-\bq)\rho_{\sigma'}(\bq)
\eeqn
where $A_{\sigma}(\bq)\equiv\left[a_{\sigma}(\bq)+
4\pi is_{\sigma}P_{\sigma}(\bq)\right]/\sqrt{8\pi s_{\sigma}}$,
$\gamma = g^*\mu_B B$ is the Zeeman energy, and
\beq
\label{coulombpotential}
v_0 = \frac{2 \pi e^2}{\epsilon q}e^{-|ql_B|^2/2}
\eeq
is the interaction potential in the lowest LL. 
The transformed density operators may be expressed in first quantization
with the help of the position of the $j$-th particle $\br_j$ and its
generalized momentum $\Pib_{\sigma}^j$,
\beq
\label{eq012}
\rho_{\sigma}(\bq)
=\sum_{j}e^{-i\bq\cdot\br_j}\left[1-\frac{il_B^2}{(1+c_\sigma)}
  \bq\times\Pib_{\sigma}^j+\mathcal{O}(q^2)\right],
\eeq
where terms of order $\mathcal{O}(q^2)$ have been omitted, and the 
oscillators are chosen to remain in their ground state, 
$\langle A_{\sigma}(\bq)\rangle=\langle A_{\sigma}^{\dagger}(\bq)\rangle=0$.
The transformed constraint is given by
\beq
\label{eq013}
\chi_\sigma(\bq)|\phi\rangle=0,
\eeq
with 
\beq
\label{eq014}
\chi_\sigma(\bq)=\sum_{j}e^{-i\bq\cdot\br_j}\left[1+il_B^2\frac{\bq\times\Pib_{\sigma}^j}
{c_\sigma(1+c_\sigma)}+\mathcal{O}(q^2)\right].
\eeq

To construct the theory at larger wave vectors, Shankar had the
ingenious insight 
that if one considers the terms in the expressions
(\ref{eq012}) and (\ref{eq014})
as the first terms of a series expansion of
an exponential,
$$\rhobar_{\sigma}(\bq)=\sum_j e^{-i\bq\cdot\bR_{\sigma,j}^{(e)}}\qquad 
{\rm and}\qquad
\chibar_{\sigma}(\bq)=\sum_j e^{-i\bq\cdot\bR_{\sigma,j}^{(v)}},$$
both operators find a compelling physical interpretation.\cite{shankar} The
operator
$$
\bR_{\sigma}^{(e)}=\br-\frac{l_B^2}{(1+c_\sigma)}\be_z\times\Pib_{\sigma}
$$
may be interpreted as the electronic {\sl guiding center} operator,
and its components satisfy the correct commutation
relations, 
$\left[X_{\sigma}^{(e)},Y_{\sigma'}^{(e)}\right]=-il_B^2\delta_{\sigma,\sigma'}$,
where the number index has been omitted.
This leads to the algebra for the Fourier components of the density in
the lowest LL \cite{GMP}
\beq
\label{eq015}
\left[\rhobar_{\sigma}(\bq),\rhobar_{\sigma'}(\bq')\right]=
\delta_{\sigma,\sigma'}2i\sin\left(\frac{\bq\wedge\bq'l_B^2}{2}\right)
\rhobar_{\sigma}(\bq+\bq'),
\eeq
where we have defined $\bq\wedge\bq'=(\bq\times\bq')_z$.
In the same manner, the components of the operator
$$
\bR_{\sigma}^{(v)}=\br+\frac{l_B^2}{c_\sigma(1+c_\sigma)}
    \be_z\times\Pib_{\sigma}
$$
satisfy $\left[X_{\sigma}^{(v)},Y_{\sigma'}^{(v)}\right]=
il_B^2\delta_{\sigma,\sigma'}/c_{\sigma}^2$, which leads to 
\beq
\label{eq016}
\left[\chibar_{\sigma}(\bq),\chibar_{\sigma'}(\bq')\right]=
-\delta_{\sigma,\sigma'}2i\sin
\left(\frac{\bq\wedge\bq' l_B^2}{2c_{\sigma}^2}\right)
\chibar_{\sigma}(\bq+\bq').
\eeq
The operator $\chibar_{\sigma}(\bq)$, which characterizes the
constraint, may thus be interpreted as the density operator of a
second type of particle with charge $-c_{\sigma}^2$, in units of the
electronic charge. However, this particle, the so-called
pseudo-vortex, only exists in the enlarged Hilbert space because it
is an excitation of the
correlated electron liquid and must be described in terms of the
electronic degrees of freedom.
A similar algebra has been investigated by Pasquier and Haldane in the 
description of bosons at $\nu=1$.\cite{PH}
The final model Hamiltonian thus reads 
\beqn
\nn
\Hhat&=&\frac{1}{2}\gamma\sum_{\sigma}\sigma \rhobar_{\sigma}(\bq=0)\\
\nn && \\ \label{eq017}
&& +\frac{1}{2}\sum_{\bq}\sum_{\sigma,\sigma'}v_0(q)    
   \rhobar_{\sigma}(-\bq)\rhobar_{\sigma'}(\bq)
\eeqn
with the constraint (\ref{eq013}) and the algebras (\ref{eq015}) and
(\ref{eq016}) for the electronic and the pseudo-vortex density,
respectively. In the case of bilayer systems, the interaction between particles in
the same layer is different (stronger) than that between particles in the different
layers. This breaks the $SU(2)$ symmetry, and one thus has to replace 
$v_0(q)\rightarrow v_0^{\sigma,\sigma'}(q)$.\cite{moon}

\subsection{Model in the CF-Basis}

In order to construct the CF representation of the model, Murthy and Shankar
proposed the ``preferred'' combination of electronic and pseudo-vortex 
density,\cite{MS,shankar}
\beq
\label{eq018}
\rhobar_{\sigma}^{P}(\bq)=\rhobar_{\sigma}(\bq)-c_{\sigma}^2
\chibar_{\sigma}(\bq),
\eeq
which plays the role of
the CF density. This approximation 
allows one to omit the constraint in the case of a gapped 
ground state,\cite{MS} whereas it fails in the limit $\nu=1/2$, where the 
system becomes 
compressible\cite{HLR} and where the constraint has to be taken into account
explicitly in a conserving approximation.\cite{approxcons}
The CF basis is  introduced after a variable transformation of the 
guiding-center coordinates $\bR_{\sigma}^{(e)}$ and $\bR_{\sigma}^{(v)}$,
\beqn
\nn
\bR_{\sigma}^{(e)}=\bR_{\sigma}+\etab_{\sigma}c_{\sigma} \qquad
\bR_{\sigma}^{(v)}=\bR_{\sigma}+\etab_{\sigma}/c_{\sigma}.
\eeqn
The new variables play the role of the CF guiding center
($\bR_{\sigma}$) and the CF cyclotron
variable ($\etab_{\sigma}$) and commute with each other. Their components, however,
satisfy the commutation relations 
$$\left[X_{\sigma},Y_{\sigma'}\right]=
-i\delta_{\sigma,\sigma'}l_{\sigma}^{*2},\qquad 
\left[\eta_{\sigma}^x,\eta_{\sigma'}^y\right]=
i\delta_{\sigma,\sigma'}l_{\sigma}^{*2},$$ 
in terms of the CF magnetic length $l_{\sigma}^*=l_B/\sqrt{1-c_{\sigma}^2}$.
The CF density operator therefore becomes 
\beq
\label{eq019}
\rhobar_{\sigma}^{P}(\bq)=\sum_{n,n';m,m'}G_{m,m'}(ql_{\sigma}^*)
F_{n,n'}(q)c_{n,m;\sigma}^{\dagger}c_{n',m';\sigma},
\eeq
where $c_{n,m;\sigma}^{\dagger}$ creates a CF with spin $\sigma$ in the state 
$|n,m\rangle$. The quantum number $n$ denotes the CF-LL,
and $m$ indicates the CF guiding-center state. The matrix 
elements are given by \cite{MS,goerbig04}
\beqn
\nn
G_{m,m'}(ql_{\sigma}^*)&=&
\sqrt{\frac{m'!}{m!}}\left(\frac{-i(q_x+iq_y)l_{\sigma}^*}{\sqrt{2}}\right)^{m-m'}
\\ \nn && \\ \label{eq020}
&&\times 
L_{m'}^{m-m'}\left(\frac{q^2l_{\sigma}^{*2}}{2}\right)e^{-q^2l_{\sigma}^{*2}/4}
\eeqn
for $m\geq m'$, and 
\beqn
\nn
F_{n,n'}(q)
&=&
\sqrt{\frac{n'!}{n!}}\left(\frac{-i(q_x-iq_y)l_{\sigma}^{*}c_{\sigma}}{\sqrt{2}}\right)^{n-n'}
\\ \nn && \\ \label{eq021} 
&&\times
  e^{-q^2l_{\sigma}^{*2}c_{\sigma}^2/4}\left[L_{n'}^{n-n'}
  \left(\frac{q^2l_{\sigma}^{*2}c_{\sigma}^2}{2}\right) \right.
\\ \nn && \\ \nn
  &&\left. 
   -c_{\sigma}^{2(1-n+n')}e^{-(ql_B)^2/2c_{\sigma}^2}
   L_{n'}^{n-n'}\left(\frac{q^2l_{\sigma}^{*2}}{2c_{\sigma}^2}\right)\right]
\eeqn
for $n\geq n'$.

We now restrict the Hilbert space to the lowest CF-LL. In this
case, the CF density operator $\rhobar^{P}_\sigma(\br)$ reads
$\langle\rhobar_{\sigma}^P(\bq)\rangle_{n=0}= 
F_{n=0}^{CF}(q)\rhobarbar_{\sigma}(\bq)$,
where 
\beq
\label{cfformfactor}
F_{n=0}^{CF}(q)=F_{0,0}(q)=e^{-|ql^*c|^2/4}
\left[1 - c^2e^{-|ql_B|^2/2c^2}\right]
\eeq
is the 
CF form factor, $p_\uparrow = p_\downarrow = 1$, $s_\uparrow =
s_\downarrow = s$, and therefore $c_\uparrow=c_\downarrow=c$ and
$l^*_\uparrow = l^*_\downarrow =l^*$ (symmetric model). The electronic filling 
factor
is thus given by $\nu=1/(2s+1)$, and the projected CF density operator is 
\beq
\label{eq021.1}
  \rhobarbar_{\sigma}(\bq)=\sum_{m,m'}G_{m,m'}(ql^*)
  c_{m,\sigma}^{\dagger}c_{m',\sigma},
\eeq
where $c_{m,\sigma}^{\dagger}$ creates a spin $\sigma$ composite
fermion in the lowest CF-LL and where we have omitted the level index.  
The projected CF density operators satisfy a similar algebra as the 
projected electron density operators,\cite{GMP,goerbig04} 
\beq
\label{eq022}
\left[\rhobarbar_{\sigma}(\bq),\rhobarbar_{\sigma'}(\bq')\right]=
\delta_{\sigma,\sigma'}2i\sin\left(\frac{\bq\wedge\bq'l_{\sigma}^{*2}}{2}
\right)
\rhobarbar_{\sigma}(\bq+\bq'),
\eeq
just with the magnetic length $l_B$ replaced by the CF magnetic length $l^*$. 
The Hamiltonian, which describes the
low-energy excitations of the FQHE system at
$\nu^*=1$ is thus given by 
\beqn
\nn
\Hhat&=&\frac{1}{2}\gamma\sum_{\sigma}
\sum_{m}\sigma c_{m,\sigma}^{\dagger}c_{m,\sigma}\\
\nn && \\ \nn
&&+\frac{1}{2}\sum_{\sigma,\sigma'}\sum_{\bq}
v_0(q)\left[F_{n=0}^{CF}(q)\right]^2
\rhobarbar_{\sigma}(-\bq)\rhobarbar_{\sigma'}(\bq) \\
\nn && \\ \label{eq023}
&=& \Hhat_Z + \Hhat_{I}
\eeqn
Note the similarity with the original Hamiltonian (\ref{eq017}); the product
$v_0(q)\left[F_{n=0}^{CF}(q)\right]^2$ plays the role of an effective 
CF interaction potential. In this sense, the FQHE system at
$\nu=1/(2s+1)$ may be treated exactly in the same manner as the  
2DEG at $\nu=1$. 

In principle, one may also describe higher CF-LLs with only one filled
spin branch. If the completely filled lower levels are assumed to
form a homogeneous
inert background, the low-energy excitations of the system at $\nu^*=2n+1$
are described by the same Hamiltonian (\ref{eq023}) now in terms of the
CF form factor of the $n$-th level, 
$F_{n}^{CF}(q)=F_{n,n}(q)$. Here,
however, we concentrate only on the case $n=0$, 
and we neglect the presence of
higher CF-LLs. This assumption becomes valid if the spin splitting is 
smaller than the typical CF-LL splitting (formally, we consider the limit
$\gamma\rightarrow 0$).

\section{Bosonization of the Model} 

As the model Hamiltonian (\ref{eq023}) for the FQHE system at
$\nu=1/(2s+1)$ resembles the one of the 2DEG at $\nu=1$, 
it will be studied with the aid
of the bosonization method for the 2DEG at $\nu=1$ recently developed 
by two of us.\cite{doretto} More precisely, within this bosonization 
approach, we can study the elementary neutral excitations 
(spin-waves, quasiparticle-quasihole pair) of the system.
Here, we will quote the main results in order to apply the
bosonization method to the case of CFs at $\nu^*=1$. 
For more details, see Ref.\ \onlinecite{doretto}. 

\begin{figure}[t]
\centerline{\includegraphics[height=2.5cm]{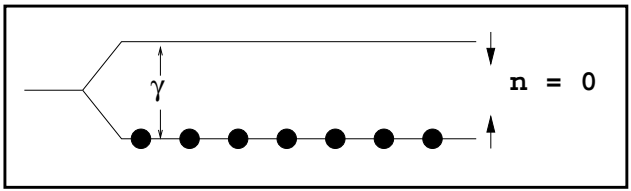}}
\caption{\label{fig01}{Schematic representation of the
    ground state of the 2DEG at $\nu^*=1$ (the quantum Hall
    ferromagnet of CFs).}}
\end{figure}

In the same way as it was done for the 2DEG at $\nu=1$, 
we first consider the noninteracting term of the Hamiltonian (\ref{eq023}),
namely the Zeeman term restricted to the lowest CF-LL, in order to
define the bosonic operators.
At $\nu^* =1$, the ground state of the noninteracting system is the
quantum Hall ferromagnet of CFs, $|FM^{CF}\rangle$,
which is illustrated in Fig.\ \ref{fig01} and which can be written as
\beq
\label{QHF}
|FM^{CF}\rangle =
 \prod_{m=0}^{N_\phi^*-1}c^{\dagger}_{m\,\uparrow}|0\rangle,
\eeq
where $|0\rangle$ is the vacuum state and $N_\phi^*$ is the
degeneracy of each CF-LL. The neutral elementary 
excitation of the system corresponds to a spin reversal within the lowest
CF-LL as one spin up CF
is destroyed and one spin down CF is created in the system. This
excited state may be built by applying the spin density operator 
$\bar{S}^- = \bar{S}_x -i\bar{S}_y$ to the ground
state (\ref{QHF}). 

We define the projected CF spin density operators in the same manner as 
for the projected CF density operators [Eq.\ (\ref{eq021.1})], 
\beqn
\bar{\bar{S}}^+_\bq &=& \sum_{m,m'}G_{m,m'}(ql^*)
           c^{\dagger}_{m\,\uparrow}c_{m'\,\downarrow}
\\ \nn && \\ 
\bar{\bar{S}}^-_\bq &=& \sum_{m,m'}G_{m,m'}(ql^*)
   c^{\dagger}_{m\,\downarrow}c_{m'\,\uparrow}
\eeqn
The commutation relation between $\bar{\bar{S}}^+_\bq$ and
$\bar{\bar{S}}^-_\bq$ may thus be written in terms of CF density operators,    
$$
[\bar{\bar{S}}^+_{\mathbf{q}},\bar{\bar{S}}^-_{\mathbf{q'}}] =
  e^{-i\bq'\wedge\bq l^{*2}/2}\rhobarbar_{\uparrow}(\mathbf{q+q'})
  -e^{-i\bq\wedge\bq' l^{*2}/2}\rhobarbar_{\downarrow}(\mathbf{q+q'}).
$$

Although the above commutation relation is quite different from the canonical
commutation relation between bosonic operators, its average value in
the ground state of the noninteracting system [Eq.\ (\ref{QHF})] is not.
Notice that, the average values of $\rhobarbar_{\uparrow}(\mathbf{q})$ and
$\rhobarbar_{\downarrow}(\mathbf{q})$ in the uniform state (\ref{QHF}) are  
$\langle\rhobarbar_{\uparrow}(\mathbf{q+q'})\rangle =
N_\phi^*\delta_{\mathbf{q+q'},0}$  and $\langle
\rhobarbar_{\downarrow}(\mathbf{q+q'})\rangle = 0$, respectively,
where $N_\phi^*=\mathcal{A}/2\pi l^{*2}$ is the number of states in one CF-LL,
in terms of the total surface $\mathcal{A}$.
Therefore, the average value of the above commutator is
\begin{equation}
\label{commutatorspin}
\langle[\bar{\bar{S}}^+_{\mathbf{q}},\bar{\bar{S}}^-_{\mathbf{q'}}]\rangle =
   N^*_{\phi}\delta_{\mathbf{q+q'},0}.
\end{equation}

With the help of the expression (\ref{commutatorspin}), we can define
the following bosonic operators 
\begin{eqnarray}
\label{b}
b_{\mathbf{q}} &=& (N_\phi^*)^{-1/2}\bar{\bar{S}}^+_{-\mathbf{q}}, \\   
&& \nonumber \\ 
\label{b+}
b^{\dagger}_{\mathbf{q}} &=& (N_\phi^*)^{-1/2}\bar{\bar{S}}^-_{\mathbf{q}}.
\end{eqnarray} 
Here, the commutation relation  
between the operators 
$b_{\mathbf{q}}$ and $ b^{\dagger}_{\mathbf{q}}$ is approximated by its
average value in the ground state (\ref{QHF}), i.e.
\begin{equation}
\label{comutadorbb}
[b_{\mathbf{q}},b^{\dagger}_{\mathbf{q'}}] \approx
\langle [b_{\mathbf{q}},b^{\dagger}_{\mathbf{q'}}] \rangle = \delta_{\mathbf{q,q'}},
\end{equation}
in analogy with Tomonaga's approach to the one-dimensional electron 
gas.\cite{tomonaga} This is the main approximation of the bosonization method. 

The quantum Hall ferromagnet of CFs $|FM^{CF}\rangle$
is the boson vacuum as the action of $c_{m'\,\downarrow}$ on
this state is equal to zero. Therefore, the eigenvectors of the
bosonic Hilbert space are
\begin{equation}
\label{bosons-eigenvalues}
|\{n_{\mathbf{q}}\}\rangle =
\prod_{\mathbf{q}}\frac{(b^{\dagger}_{\mathbf{q}})^{n_{\mathbf{q}}}}
  {\sqrt{n_{\mathbf{q}}!}}|FM^{CF}\rangle,
\end{equation}
with $n_\bq > 0$ and $\sum n_\bq < N_\phi^*$. In addition, note
that the state $b^\dagger_\bq|FM^{CF}\rangle$ is a linear combination
of particle-hole excitations.

One also finds the bosonic representation of the projected CF density 
operators $\rhobarbar_{\sigma}(\mathbf{q})$,\cite{doretto}
\begin{equation}
\label{densityoperator2}
 \rhobarbar_{\sigma}(\mathbf{q}) = \delta_{\sigma,\uparrow}\delta_{\mathbf{q},0}N^*_{\phi}
 - \sigma\sum_{\mathbf{k}}
  e^{-\sigma i\mathbf{q}\wedge\mathbf{k}l^{*2}/2}b^{\dagger}_{\mathbf{q}+\mathbf{k}}b_{\mathbf{k}}.
\end{equation}
One can see that the CF density operator is {\it quadratic} in the bosonic operators 
$b$.
In contrast to the expressions of Ref.\ \onlinecite{doretto},
the Gaussian factor $e^{-|ql_B|^2/4}$ is absent in Eqs.\ (\ref{b}),
(\ref{b+}) and (\ref{densityoperator2}) as it is included in the  
interaction potential $v_0(q)$
[Eq.\ (\ref{coulombpotential})]. Therefore, from
Eq.\ (\ref{densityoperator2}),
the CF density operator
$\rhobarbar(\bq)$ has the following bosonic representation
\begin{eqnarray}
\rhobarbar(\bq) 
&=& \rhobarbar_{\uparrow}(\mathbf{q}) + \rhobarbar_{\downarrow}(\mathbf{q})
\label{densityoperator}  \\
&& \nonumber \\ \nonumber
&=& \delta_{\mathbf{q},0}N_{\phi}^* + 2i\sum_{\mathbf{k}}
\sin(\mathbf{q}\wedge\mathbf{k}l^{*2}/2)b^{\dagger}_{\mathbf{q}+\mathbf{k}}b_{\mathbf{k}}.
\end{eqnarray}

We are now able to study the complete Hamiltonian
(\ref{eq023}), i.e., the interacting system at $\nu^*=1$. 
The bosonic representation of the Zeeman term $\Hhat_Z$ is given by\cite{doretto}   
\begin{equation}
\label{zeemanboson}
\hat{H}_Z = \gamma\sum_{\mathbf{q}}b^{\dagger}_{\mathbf{q}}b_{\mathbf{q}}
      - \frac{1}{2}\gamma N^*_{\phi}.
\end{equation}
Because
the CF density operator is {\it quadratic} in the bosonic operators, 
$\Hhat_{I}$ in Eq. (\ref{eq023}) is {\it quartic} in the operators $b$, and one
obtains with the help of Eq. (\ref{densityoperator})
\begin{eqnarray}
\nonumber
\Hhat_{I} &=& \sum_{\mathbf{p,q}}\int \frac{d^2k}{2\pi^2}\;
v_0(\mathbf{k}) 
        \left[F_{0}^{CF}(k)\right]^2\sin(\mathbf{k}\wedge\mathbf{p}l^{*2}/2)
\\ \nonumber 
&& \\ \nonumber
        &&\times\sin(\mathbf{k}\wedge\mathbf{q}l^{*2}/2)
        b^{\dagger}_{\mathbf{k}+\mathbf{q}}b_{\mathbf{q}}
                          b^{\dagger}_{\mathbf{p}-\mathbf{k}}b_{\mathbf{p}} \\ \nonumber
 &&\\\nonumber
    &=& \sum_{\mathbf{q}}\int
    \frac{d^2k}{2\pi^2}\;v_0(\mathbf{k})\left[F_{0}^{CF}(k)\right]^2 
        \sin^2(\mathbf{k}\wedge\mathbf{q}l^{*2}/2)
        b^{\dagger}_{\mathbf{q}}b_{\mathbf{q}} \\ \nonumber 
 &&\\\nonumber
    && + \sum_{\mathbf{p,q}}\int
    \frac{d^2k}{2\pi^2}\;v_0(\mathbf{k})\left[F_{0}^{CF}(k)\right]^2 
       \sin(\mathbf{k}\wedge\mathbf{p}l^{*2}/2)
\\ \nonumber && \\ \label{vboson}
&&     \times\sin(\mathbf{k}\wedge\mathbf{q}l^{*2}/2)
       b^{\dagger}_{\mathbf{k}+\mathbf{q}} b^{\dagger}_{\mathbf{p}-\mathbf{k}}
                         b_{\mathbf{q}}b_{\mathbf{p}}. \\ \nonumber
\end{eqnarray}
Adding Eq.\ (\ref{zeemanboson}) to (\ref{vboson}) yields the
expression for the total Hamiltonian of the FQHE system at $\nu^*=1$ 
in terms of bosonic operators,
\beq\label{hintboso}
\Hhat^B = \Hhat^B_0 + \Hhat^B_{int},
\eeq
with the free boson Hamiltonian 
\beq
\Hhat^B_0 = -\frac{1}{2}\gamma N^*_{\phi} +
   \sum_{\mathbf{q}}w_{\mathbf{q}}b^{\dagger}_{\mathbf{q}}b_{\mathbf{q}}
\eeq
and the boson-boson interaction part
\beqn
\nn
\Hhat^B_{int} &=& 
\frac{2}{\mathcal{A}}\sum_{\mathbf{k,p,q}}v_0(k)\left[F_{0}^{CF}(k)\right]^2
       \sin(\mathbf{k}\wedge\mathbf{p}l^{*2}/2) \\ 
    && \times\sin(\mathbf{k}\wedge\mathbf{q}l^{*2}/2)
      b^{\dagger}_{\mathbf{k}+\mathbf{q}} b^{\dagger}_{\mathbf{p}-\mathbf{k}}
                         b_{\mathbf{q}}b_{\mathbf{p}}.
\eeqn
The dispersion relation of the bosons is given by
\begin{equation}
\label{rpa}
w_{\mathbf{q}} = \gamma + \int \frac{d^2k}{2\pi^2}\;v_0(\mathbf{k})\left[F_{0}^{CF}(k)\right]^2
        \sin^2(\mathbf{k}\wedge\mathbf{q}l^{*2}/2).  
\end{equation}

\begin{figure}
\epsfysize+4.5cm
\epsffile{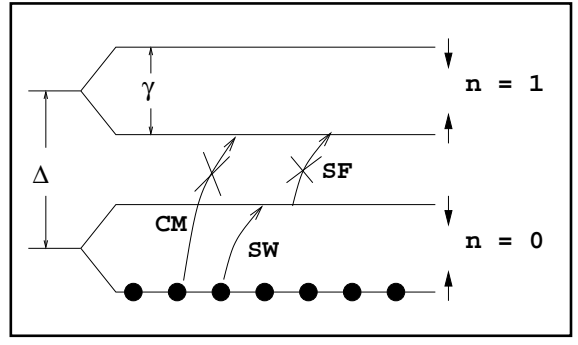}
\caption{Schematic representation of the neutral excitations of the 
         quantum Hall ferromagnet of CFs at $\nu^*=1$. The
         bosonization method takes into account only spin-wave (SW) excitations 
         which involve a spin reversal inside the lowest CF-LL. The
         excitations between different CF-LL, without a spin reversal
         (charge mode, CM) and with a spin-flip (SF) are neglected.}
\label{fig02}
\end{figure}

Therefore, the bosonization method for the 2DEG at $\nu = 1$ allows us
to map the fermionic model Hamiltonian (\ref{eq023}) for the
FQHE system at $\nu^* = 1$,
into an interacting bosonic one. 
The ground state of the bosonic model (\ref{hintboso}) is the boson vacuum,
which corresponds to the quantum Hall ferromagnet of
CFs, with energy $E_{FM} = -\gamma N_\phi^*/2$. In fact, this
is also the ground state of the noninteracting
bosonic model (\ref{zeemanboson}), and thus the ground state of
the system does not change when the interaction between the electrons is
introduced. 

The bosonization approach \cite{doretto}
restricts the total Hamiltonian (\ref{eq023}) to one particular sector
of all excitations of the quantum Hall ferromagnet of CFs, namely the
one which corresponds to a spin reversal inside the lowest CF-LL:
a spin-wave (SW) excitation (Fig.\ \ref{fig02}). 
It neglects excitations between
different CF-LLs, both those without spin reversal (charge mode,
CM) and those which also involve a spin-flip (SF).

A further issue in the present bosonization scheme would be to provide
the expression of the fermion field operator in terms of the bosonic ones,
$b^\dagger_{\mathbf{q}}$ and $b_{\mathbf{q}}$, i.e., the analog of the
Mattis-Mandelstam formula for one-dimensional fermionic systems.
\cite{delf,voit} Such expression would enable one to calculate, for
instance, correlation functions related to the model (41). Here, however,
we are interested only in the energy dispersion of the collective 
spin excitations, which find their natural formulation in terms of the bosonic
operators and which may be calculated without the precise expression for the
fermionic fields.

\subsection{Dispersion relation of the bosons}  

In this section, we evaluate the dispersion relation of the
bosons [Eq. (\ref{rpa})] and investigate whether the interaction
potential between them accounts for the formation of a bound state of
two bosons, as it was found for the 2DEG at $\nu=1$.
In addition to the bare Coulomb potential, we 
also consider the modification of the interaction potential between
the original fermions due to finite-width effects.

\begin{figure}[t]
\centerline{\includegraphics[height=10.0cm]{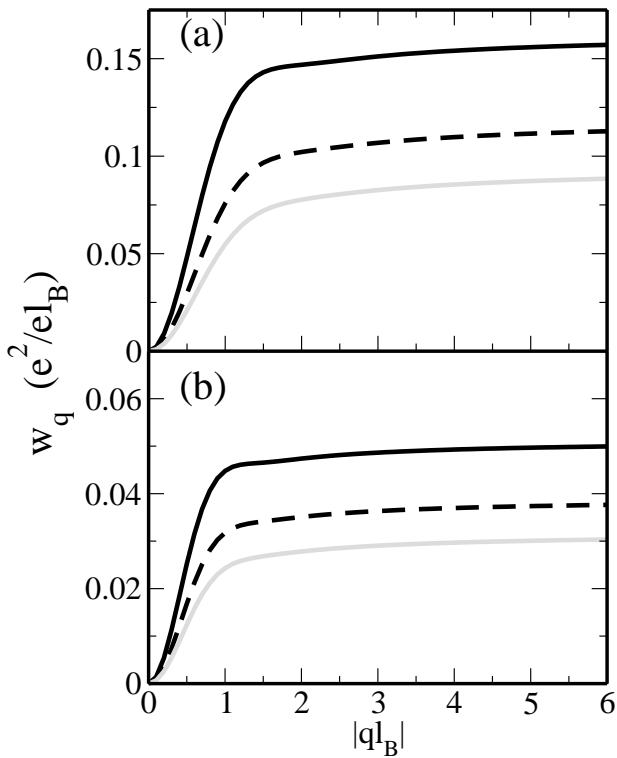}}
\caption{\label{fig03}{Dispersion relations of the bosons as a
  function of the momentum $\bq$ with $\gamma=0$, in units of
  $e^2/\epsilon l_B$, when (a) $\nu=1/3$ and (b)
  $\nu=1/5$.  
  The solid line corresponds to Eq.\ (\ref{rpacoulomb}), whereas the
  dashed and solid gray lines represent the results for the width
  parameters $\lambda = 0.5$ and $1.0$, respectively.}}
\end{figure}

For the Coulomb potential [Eq. (\ref{coulombpotential})], the integral 
over momentum in (\ref{rpa}) may be performed analytically,
\begin{widetext}
\beqn
\nn
w_{\mathbf{q}} 
       &=& \gamma +\int \frac{d^2k}{2\pi^2} \;v_0(\mathbf{k})
               e^{-|kl^*c|^2/2}
                 \left[1 - c^2e^{-|kl_B|^2/2c^2}\right]^2
                 \sin^2(\mathbf{k}\wedge\mathbf{q}l^{*2}/2)
\\ \nn && \\ \nn
       &=& \gamma + \frac{e^2}{\epsilon l_B}\sqrt{1-c^2}\sqrt{\frac{\pi}{2}}
               \left[\alpha(c) -  e^{-|ql^*|^2/4}
{\rm I_0}\left(\frac{|ql^*|^2}{4}\right)\right.
             - \frac{c^5}{\sqrt{2-c^2}}e^{-|ql^*c|^2/4(2-c^2)}
             {\rm I_0}\left(\frac{|ql^*c|^2}{4(2-c^2)}\right)
\\ \nn && \\ \label{rpacoulomb}              
       &&  \left. + 2c^3e^{-|ql^*c|^2/4}
             {\rm I_0}\left(\frac{|ql^*c|^2}{4}\right)\right],
\eeqn
\end{widetext} 
where ${\rm I_0(x)}$ is the modified Bessel function of the first kind.\cite{arfken}
The function  
$$
\alpha(c) = 1 - 2c^3 + \frac{c^5}{\sqrt{2-c^2}}
$$
depends on the charge of the pseudo-vortex,
$-c^2$, which is related to the electronic filling factor
$\nu$. Remember that, for $\nu = 1/(2s + 1)$ we have $c^2=2s/(2s+1)$. For the
case of vanishing Zeeman energy ($\gamma=0$),
the results of Eq.\ (\ref{rpacoulomb}) are shown in Figs.\
\ref{fig03}\ (a) and (b)
(solid lines) for $\nu=1/3$ and $1/5$, respectively.
The energies are given in units of $e^2/\epsilon l_B$.

Notice that at $q=0$ the energy of the bosons is equal to the Zeeman
energy $\gamma$ (in agreement with Goldstone's theorem) and that $w_q \sim
|ql_B|^2$ in the small-wavevector limit, $|ql_B| \ll 1$ [see Eq.\ (\ref{rpacoulombapprox})].
This is exactly the behavior of spin-wave excitations in a ferromagnetic
system. Therefore, the long-wavelength limit of
$w_\bq$ corresponds to spin-wave excitations of the quantum Hall
ferromagnet of CFs. To the best of our knowledge, this is the first analytical
calculation of $w_\bq$.

In analogy with the quantum Hall ferromagnet at $\nu = 1$, we may 
associate the limit $q \rightarrow \infty$ of $w_\bq$
to the energy $E_{ph}$ to create a well-separated
CF-quasiparticle-quasihole pair, i.e.
\beqn
\label{qp-qh}
E_{ph} &=& \gamma + \sqrt{\frac{\pi}{2}}\sqrt{1-c^2}\alpha(c)\frac{e^2}{\epsilon l_B} 
        = \gamma + E^C_{ph},
\eeqn    
where 
\beq
\label{MS-qp-qh}
E^C_{ph} = \left\{\begin{array}{r@{\quad}l} 
                0.163 & \nu=1/3, \\
                0.051 & \nu=1/5,
               \end{array} \right.
\eeq
in units of $e^2/\epsilon l_B$.

Our result is in agreement with the one found by Murthy in the
``naive'' approximation, where the dispersion relation 
of spin-wave excitations of the FQHE system
(neglecting disorder, finite-width and Landau-level-mixing effects) at
$\nu=1/3$ has been evaluated using the time-dependent
Hartree-Fock approximation.\cite{murthy}   
It is not a surprise that both calculations are in agreement as the main
approximation of our bosonization method [Eq.(\ref{commutatorspin})]
resembles the one considered in the random-phase approximation 
(RPA).\cite{pines}

On the other hand, our results appear to be overestimated
when compared to the ones of Nakajima and Aoki, who have
also calculated the spin wave excitation of the FQHE system at $\nu=1/3$ and
$1/5$.\cite{nakajima} They started from the results of Kallin and Halperin
for the 2DEG at $\nu = 1$,\cite{kallin} turned to the spherical-geometry
approach, introduced the CF picture and then performed  
mean-field calculations. Their results agree with the
ones of Ref. \onlinecite{kallin} for the 2DEG at $\nu = 1$ with the
replacements $e \rightarrow e^*$ and $l_B \rightarrow l^*$.   
Finally, Mandal has studied 
spin-wave excitations in the
framework of a fermionic Chern-Simons gauge theory.\cite{mandal} 
For $\nu = 1/(2s+1)$, his results coincide with those obtained by
Kallin and Halperin for the 2DEG at $\nu = 1$, with the replacement
$l_B \rightarrow l^*$, wheras the charge remains unchanged. Mandal's estimates
for the energy of a well separated CF quasiparticle-quasihole pair are 
therefore larger in comparison with the above mentionned works.
A possible reason for this mismatch may be the presence of higher CF-LLs, 
which are implicitely taken into account in the numerical approach 
by Nakajima and Aoki\cite{nakajima} and explicitely in Murthy's time-dependent
Hartree-Fock calculations, beyond the naive approximation.\cite{murthy} 
Remember that the model (\ref{eq023}) used in our analyses, however, is 
restricted to a {\sl single} level and thus ignores the
presence of higher CF-LLs. 
The latter may in principle be included in a 
screened interaction potential, the $q$-dependent dielectric function of 
which may be calculated within the RPA.\cite{goerbig04} 
However, this would only take into account virtual density-density 
fluctuations, but no single-particle excitations. 

The energy $E^C_{ph}$ has been calculated numerically by Mandal and Jain
for the FQHE system 
(neglecting disorder, finite width and Landau level mixing)
at $\nu=1/3$, using trial wavefunctions for
the ground state in the spherical-geometry
approach.\cite{mandal-jain}
The thermodynamic limit obtained from the calculations on finite
systems indicated that $E^C_{ph} = 0.074\;e^2/\epsilon l_B$, in
agreement with the previous results by Rezayi,\cite{rezayi2} 
who performed numerical-diagonalization calculations on finite
systems, {\it without} the CF picture. Notice that this value is almost
half as large as the one we obtain in the framework of the Hamiltonian
theory [Eq.\ (\ref{MS-qp-qh})]. 

In order to include finite-width effects, we
replace the bare Coulomb potential (\ref{coulombpotential}) by 
$$
v_0^{\lambda}(q) = \frac{2\pi e^2}{\epsilon q}e^{-|ql_B|^2/2}
          e^{|q\lambda|^2}\left[1 - {\rm Erf}(\lambda q)\right],
$$
where ${\rm Erf}(x)$ is the error function.\cite{arfken} This expression
has been obtained under the assumption that the confining potential in
the $z$-direction (with a characteristic width $\lambda$, in units of $l_B$) is
quadratic. \cite{MS,shankar,morf} 
After replacing $v_0(q)\rightarrow v_0^{\lambda}(q)$, the momentum integral in
Eq. (\ref{rpa}) is evaluated numerically. For $\nu=1/3$ and $1/5$, the results for
two different values of the width parameter ($\lambda=0.5$ and $1.0$, dashed and
gray lines, respectively) are shown in Figs. \ref{fig03}\ (a) and (b).

Notice that, in agreement with results by Murthy, the energy of the bosons
decreases as $\lambda$ increases.\cite{murthy} In fact, the finite-width
results are likely to be more reliable than the ones at $\lambda = 0$ 
as the preferred combination for the 
charge density [Eq.\ (\ref{eq021})] is derived in the long-wavelength
limit, and the inclusion of finite-width effects cuts off
the short-range contributions of the Coulomb potential.   

From the experimental point of view, spin reversal excitations of the
FQHE system have been studied with the help of inelastic light 
scattering.\cite{kang,dujvone1,dujvone2} In this case, it is possible to
probe some critical points of the dispersion relation of the neutral
excitation, namely the ones at $q\rightarrow 0$ and
$q\rightarrow \infty$. In particular, the latter can be probed because
the wavevector is no longer conserved in the presence of 
residual disorder. The experimental spectrum at $\nu=1/3$ consists of five 
different peaks.\cite{dujvone1,dujvone2}  
Three of them have been associated to the charge
modes at $q\rightarrow 0$ ($\Delta_0$), $q\rightarrow \infty$
($\Delta_\infty$), and to the magnetoroton minimum. The one at the
Zeeman energy has been associated to the long-wavelength limit of the spin-wave
excitations, while the one between $\Delta_0$ and $\Delta_\infty$ was
related to $E_{ph}$. In this case, it was estimated that
$E_{ph}^C = 0.054\;e^2/\epsilon l_B$. 
We obtain precisely this value for a width parameter $\lambda=3.0$, which 
corresponds to $290\,$\AA ~at $B=7\,$T ($l_B=97\,$\AA). 
In view of the approximations made (no Zeeman splitting,
restriction to a single CF-LL level, and neglecting impurity effects),
this is surprisingly
close to the experimental situation, where
a quantum well of width $330\,$\AA ~was investigated at 
$7\,$T.\cite{dujvone1,dujvone2}

\subsection{\label{sec:twobosons} Bound States of Two Bosons}  

Apart from the fact that our bosonization scheme
reproduces the time-dependent Hartree-Fock (or RPA) results, it goes
beyond as it yields an {\sl interacting} bosonic model
(\ref{hintboso}). Such kind of result was first derived for the 2DEG at
$\nu=1$ in Ref.\ \onlinecite{doretto}. In this section, we follow the
lines of this work and study two-boson states in the
interacting bosonic model in order to find out a relation
between the two-boson bound states and the 
quasiparticle-quasihole (small skyrmion-antiskyrmion) pair excitations.

Finite-width effects are not considered here, and we therefore
study the bosonic model (\ref{hintboso}) with the dispersion relation
of the bosons given by Eq.\ (\ref{rpacoulomb}). 
In order to study two-boson states, one needs to solve the
Schr\"odinger equation
\begin{equation}
\mathcal{H}|\Phi_{\mathbf{P}}\rangle = E_{\mathbf{P}} 
|\Phi_{\mathbf{P}}\rangle,
\label{eqsch}
\end{equation}
where $|\Phi_{\mathbf{P}}\rangle$ is the more general representation of a 
two-boson state with total momentum $\bP$, namely 

\begin{equation}
|\Phi_{\mathbf{P}}\rangle = \sum_{\mathbf{q}}\Phi_{\mathbf{P}}(\mathbf{q})
                            b^{\dagger}_{\mathbf{\frac{P}{2}-q}}
                            b^{\dagger}_{\mathbf{\frac{P}{2}+q}}|FM\rangle.
\label{2bosons}
\end{equation}
Here, $\bP$ is a good quantum number because the total momentum of a
2D system of charged particles in an external magnetic
field is conserved when the total charge of the two-particle state is
zero.\cite{osborne}    

Substituting Eq.\ (\ref{2bosons}) into (\ref{eqsch}) we obtain the 
following eigenvalue equation 
\begin{equation}
[\varepsilon - E_{\mathbf{P}}(\mathbf{q})]\Phi_{\mathbf{P}}(\mathbf{q})
= \int d^2k\, K_{\mathbf{P}}(\mathbf{k-q,q})\Phi_{\mathbf{P}}(\mathbf{k}),
\label{autovalores}
\end{equation}
where $E_\bP(\bq) = w_{\bP/2 - \bq} + w_{\bP/2 + \bq}$ is the energy of
two free bosons, $\varepsilon = E_{\mathbf{P}} - E_{FM}$, and the kernel of the
integral equation reads  
\begin{eqnarray}
K_{\mathbf{P}}(\mathbf{k-q,q}) &=& \frac{2}{\pi}\frac{e^2}{\epsilon l_B}
       \frac{e^{-|\mathbf{k-q}|^2/2(1-c^2)}}{|\mathbf{k-q}|}
       \left[F^{CF}_0(|\bk-\bq|)\right]^2
\nonumber \\ && \nonumber \\
  &&   \times\sin\left(\frac{(\mathbf{k-q})\wedge(\mathbf{P}/2+\mathbf{q})}{2(1-c^2)}\right)
\nonumber \\ && \nonumber \\
  && \times\sin\left(\frac{(\mathbf{k-q})\wedge(\mathbf{P}/2-\mathbf{q})}{2(1-c^2)}\right).
\;\;\;\;\;
\label{kernel}
\end{eqnarray}
In the above expressions, all momenta are measured in units of the
inverse of the magnetic length $l_B^{-1}$.

The eigenvalues of (\ref{autovalores})
are evaluated numerically, using the quadrature
technique.\cite{arfken}  
In this case, the integral over momenta in
(\ref{autovalores}) is replaced by a set of algebraic equations,
\begin{equation}
[\varepsilon - E_{\mathbf{P}}(\mathbf{q_i})]\Phi_{\mathbf{P}}(\mathbf{q_i})
\approx \sum_{j\not=i} C_jK_{\mathbf{P}}(\mathbf{q_j-q_i,q_i})
\Phi_{\mathbf{P}}(\mathbf{q_j}),
\label{autovalores1}
\end{equation}
and the eigenvalues for a fixed momentum $\bP$ may be
calculated by means of usual matrix techniques. The coefficients of
the quadrature $C_j$ and the points $\bq_j$ depend on the
parametrization. We chose the one with $61$ points, which is used
to calculate two-dimensional integrals over a circular region.\cite{stroud}  
A cut-off for large momenta has to be introduced in
order to define the region of integration. Therefore, only bosons 
with $|ql_B| < 1$ 
are included in our calculations, and we concentrate 
on the analysis of the lowest-energy two-boson state. 
Both restrictions are related to the limitations of the
numerical method.

\begin{figure}[t]
\centerline{\includegraphics[height=10.0cm]{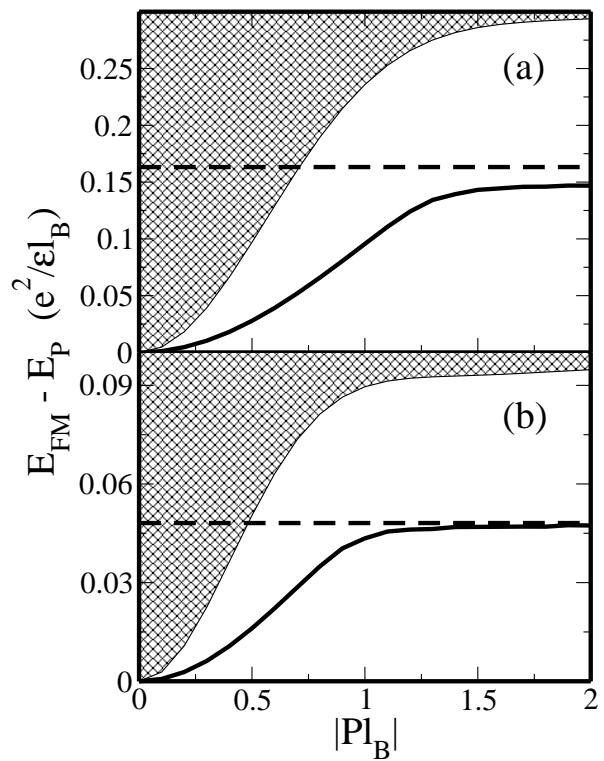}}
\caption{\label{fig05}{Dispersion relations of the two-boson state as a
  function of the total momentum $\bP$, with $\gamma=0$, in units of
  $e^2/\epsilon l_B$, when (a) $\nu=1/3$ and (b) $1/5$. 
  The solid line corresponds to the lowest-energy state of two bosons,
  the dashed line is the
  energy of a well-separated CF quasiparticle-quasihole
  [Eq.\ (\ref{qp-qh})], and 
  the shaded area represents the continuum of scattering states.}}
\end{figure}

In Figs.\ \ref{fig05}\ (a) and (b), we show the lowest-energy
state of two bosons (solid line) as a function of the total momentum $\bP$ for
$\nu=1/3$ and $1/5$, respectively, in units of 
$e^2/\epsilon l_B$, in the limit of $\gamma =0$. The shaded area corresponds to
the continuum of scattering states. Because the solid line is below the 
band of scattering states, we may interpret the lowest energy eigenvalue 
($E_\bP$) as a bound state of two bosons.
We have also included (dashed line) the value of the energy $E^C_{ph}$ of a 
CF-quasiparticle-quasihole pair [Eq.\ (\ref{qp-qh})], with the particle and
the hole far apart from each other. Notice that in both
cases, the asymptotic limit ($|\bP| \rightarrow \infty$) of
$E_\bP$ is close to $E^C_{ph}$. 
In analogy with the 2DEG at $\nu=1$,\cite{doretto}
the bound state of two bosons with
large $\bP$ may be understood, for instance, as
a CF quasiparticle bound to one spin wave plus a distant
CF quasihole.\cite{doretto} In this scenario, our results indicate that
the creation of a CF quasiparticle dressed by a spin wave (or alternatively,
a small skyrmion of CFs) is more favorable for the FQHE system at
$\nu=1/3$ than for the one at $\nu=1/5$ as $|E^C_{ph} -
E_{P\rightarrow\infty}|$ is smaller in the latter case.    
At present, no Knight-shift data are available for FQHE systems at
$\nu=1/5$ to check our theoretical results. 

The energy of small CF skyrmions 
for FQHE systems at $\nu=1/3$ and $1/5$ has been calculated 
numerically by Kamilla {\it et al.},\cite{kamilla} using 
a hard-core wavefunction.\cite{macdonald}
However, it is not possible to compare our results, namely, the
asymptotic limit of $E_\bP$, to the ones obtained by Kamilla and co-workers as
they have evaluated the energy {\it only} of the positively charged
excitation. In contrast to the 2DEG at $\nu=1$, 
the particle-hole symmetry is broken at $\nu=1/3$ and $1/5$.
Therefore, the energy of the positive excitation cannot be used to
estimate the energy of the negative one, and consequently the
energy of a well-separated skyrmion-antiskyrmion pair cannot be determined.
The breaking of the particle-hole symmetry at $\nu=1/3$ may be observed in 
the Knight-shift data, which are asymmetric around this filling 
factor.\cite{khandelwal2}

\section{Skyrmions of CFs} 

It has been shown in Ref.\ \onlinecite{doretto} that the semi-classical limit 
of the 
interacting bosonic model provides a microscopic basis for a
phenomenological model used by Sondhi {\sl et al.}\cite{sondhi} for the description
of quantum Hall skyrmions. 
Sondhi's model is given by the Lagrangian density  
\begin{eqnarray}
\nonumber
\mathcal{L}_{eff} &=&
 \frac{1}{2}\rho_0\mathcal{A}(\mathbf{n})\cdot\partial_t\mathbf{n}
 -\frac{1}{2}\rho_S(\nablab\mathbf{n})^2
 +\frac{1}{2}\gamma\rho_0\mu_B\mathbf{n}\cdot\mathbf{B}
\\ \nonumber && \\ \label{lagsondhi}
 && -\frac{e^2}{2\epsilon}\int
 d^2r'\frac{q(\mathbf{r})q(\mathbf{r}')}{|\mathbf{r}-\mathbf{r}'|}, 
\end{eqnarray}
where $\rho_S$ is the spin stiffness,\cite{comment1}
$\mathcal{A}(\mathbf{n})$ is the vector potential 
of a unit monopole ($\epsilon^{abc}\partial_a\mathcal{A}_b = n^c$)
and $\rho_0 = \nu/(2\pi l_B^2)$ is the average electronic density. 
The topological charge or skyrmion density is given by 
\begin{equation}
q(\mathbf{r}) = \frac{1}{8\pi}\epsilon^{\alpha\beta}\mathbf{n}\cdot
  (\partial_{\alpha}\mathbf{n}\times\partial_{\beta}\mathbf{n}),
\label{cargatopologica}
\end{equation}
with $a,b,c = x,y,z$ and $\alpha,\beta = x,y$.
Within the bosonization approach, the
gradient, Zeeman and interaction terms of the above Lagrangian density
have been derived, and one obtains, from a microscopic description, exactly 
the same coefficients.  
This result corroborates the relation
between the bound state of two bosons and the small
skyrmion-antiskyrmion pair excitation discussed in
Ref.\ \onlinecite{doretto}. In other words, the interacting bosonic
model may be used to describe the quantum Hall skyrmion at $\nu=1$.

Here, we proceed in the same manner in order to derive the analogue of 
Sondhi's model for the CF skyrmion 
at $\nu=1/(2s+1)$. 
We analyze the bosonic model (\ref{hintboso}), with the
dispersion relation of the bosons given by Eq.\ (\ref{rpacoulomb}). 

Following the lines of Ref.\ \onlinecite{moon}, we 
describe the CF skyrmion by the coherent
state 
\begin{equation}
\label{textura}
   |sk^{CF}(\mathbf{n})\rangle = \exp(-i\hat{O})|FM\rangle,
\end{equation}
where the operator $\hat{O}$ is a non-uniform spin rotation which
reorients the local spin from the direction $\mathbf{e}_z$ to
$\mathbf{n}(\mathbf{r})$,
\begin{eqnarray}
\nonumber
\hat{O} &=& \int d^2r\; \mathbf{\Omega}(\mathbf{r})\cdot\mathbf{S}(\mathbf{r})
\\ \nonumber  &&\\ \label{opO}
            &=& \int d^2q\; \left[\Omega^+(\mathbf{q})S^-_{-\mathbf{q}}
                       + \Omega^-(\mathbf{q})S^+_{-\mathbf{q}}\right].
\end{eqnarray}
Here, $\mathbf{S}(\mathbf{r})$ is the spin operator,
$\mathbf{n}(\mathbf{r})$ is a unit vector, and
$\mathbf{\Omega}(\mathbf{r}) = \mathbf{e}_z\times\mathbf{n}(\mathbf{r})$
defines the rotation angle. We assume that  $\mathbf{\Omega(\mathbf{r})}$
describes small tilts away from the $\mathbf{e}_z$ direction, i.e.
$\Omega^{\pm}(\mathbf{q})$ vanishes when $|ql_B|\gg 1$.

Using the expression for the spin density operator in terms of the
boson operator $b$,\cite{comment2} 
the operator $\hat{O}$ becomes
\begin{equation}
\hat{O} \equiv \beta \sum_{\bq}
             \left[\Omega^+_{\mathbf{q}}b^{\dagger}_{-\mathbf{q}}
                   + \Omega^-_{\bq}b_\bq\right],
\end{equation}
and the constant $\beta$ will be defined later. 

We start with the analysis of the CF density operator
(\ref{densityoperator}). The average value of this operator in the
state (\ref{textura}) is
\begin{eqnarray}
\nonumber
\langle \rhobarbar(\bk)\rangle &=&
    2i\beta F^0_{CF}(k)
\\ \nonumber && \\ \label{averagedensity1} 
  &&\times\sum_\bq\sin\left(\mathbf{k}\wedge\mathbf{q}l^{*2}/2\right)
    \Omega^-_{\bk + \bq}\Omega^+_{-\bq}.
\end{eqnarray}
Because only the long-wavelength limit of the theory is considered, one
may approximate
$$
   \sin\left(\mathbf{k}\wedge\mathbf{q}l^{*2}/2\right) \approx
   \mathbf{k}\wedge\mathbf{q}l^{*2}/2 =
   (\mathbf{k}\times\mathbf{q})_z l^{*2}/2,
$$
and the CF form factor $F^0_{CF}(k) \approx 1 - c^2$.

This implies that we only retain terms $\mathcal{O}(k^2)$ in
Eq.\ (\ref{averagedensity1}). Replacing the sum over momenta by an
integral, and using the relation between the vector
$\mathbf{\Omega}(\br)$ and the unit vector field $\mathbf{n}(\br)$,
the average value of the CF density may be written as 
\begin{widetext}
\begin{eqnarray}
\langle \rhobarbar(\bk)\rangle
  &=& (1 - c^2)\frac{iN^*_{\phi}\beta^2(l^*)^4}{2\pi}
      \int d^2q\;\mathbf{e}_z\cdot\left[(\mathbf{k}+\mathbf{q})\Omega^-_{\mathbf{k}+\mathbf{q}}
      \times\mathbf{q}\Omega^+_{\mathbf{q}}\right] \nonumber \\
\label{averagedensity2}
  &=& -(1 - c^2)2\pi N^*_\phi \beta^2(l^*)^4
      \int d^2r\;e^{-i\mathbf{k}\cdot\mathbf{r}}
\mathbf{e}_z\cdot\left(\nablab n^x\times\nablab n^y-\nablab n^y\times\nablab n^x\right),
\end{eqnarray}
and therefore, its Fourier transform becomes
\begin{equation}
\label{skyrmioncharge}
\langle \rhobarbar(\br)\rangle
       = -(1 - c^2)2\pi N^*_{\phi}\beta^2(l^*)^4  
\mathbf{e}_z\cdot\left(\nablab n^x\times\nablab n^y-\nablab n^y\times\nablab n^x\right).
\end{equation}
\end{widetext}
Notice that, if we choose the constant 
$\beta = 1/\left( 4\pi\sqrt{N^*_\phi}(l^*)^2\right)$
the average value of the CF density is equal to the topological charge
times the filling factor $\nu = (1-c^2) = 1/(2s+1)$, as suggested by
Sondhi {\it et al.} in Ref.\ \onlinecite{sondhi}. 
With this choice, we may calculate
the average value of the energy in the state (\ref{textura}), considering
the interacting bosonic model (\ref{hintboso}), with $w_\bq$ as in
Eq.\ (\ref{rpacoulomb}).   

The average value of the quadratic term of the Hamiltonian
(\ref{hintboso}) is given by
\begin{eqnarray}
\nonumber
\langle sk^{CF}|\Hhat_0^B|sk^{CF}\rangle
  &=&\frac{1}{16\pi^2N^*_\phi (l^*)^4}\sum_\bq
     w_{\mathbf{q}}\Omega^+_{-\mathbf{q}}\Omega^-_{\mathbf{q}}.
\end{eqnarray}
Because the long-wavelength limit of the dispersion relation of the bosons
[Eq.\ (\ref{rpacoulomb})] is 
\beq
\label{rpacoulombapprox}
w_\bq \approx \gamma + \frac{e^2}{l_B\epsilon}\sqrt{1-c^2}\sqrt{\frac{\pi}{2}}
                  \frac{1}{4}\tilde{\alpha}(c)|ql^*|^2,
\eeq
with the constant
\beq
\tilde{\alpha}(c) = 1 - 2c^5 + \frac{c^7}{(2-c^2)^{3/2}},
\eeq
the average value of $\Hhat_0^B$ is
\begin{eqnarray}
\nonumber
\langle \mathcal{H}_0\rangle
    &\approx& \frac{1}{2}\rho_S\int d^2r\;
              \left[\nablab\mathbf{n}(\mathbf{r})\right]^2
\\ \nonumber && \\ \nn
             && - \frac{1}{2}g^*\mu_B\frac{1}{2\pi (l^*)^2}
              \int d^2r\;\mathbf{n}\cdot\mathbf{B}
\\ \nonumber && \\ \label{energyskyrmion1}
             && + \frac{1}{2}g^*\mu_BB\frac{1}{2\pi}N^*_\phi.
\end{eqnarray}
Here, the spin stiffness is defined as 
\beq
\label{stiffness}
\rho_S = \frac{1}{16\pi}\frac{e^2}{\epsilon l^*}\sqrt{\frac{\pi}{2}}
            \tilde{\alpha}(c)=\rho_S^*\mu(c),
\eeq
with
\beq
\label{stiffness2}
\rho_S^* \equiv \frac{1}{16\pi}\frac{e^{*2}}{\epsilon l^*}\sqrt{\frac{\pi}{2}}
\eeq
and $\mu(c)=\tilde{\alpha}(c)/(1-c^2)^2$. Notice that one would have obtained
the value $\rho_S^*$ for the CF spin stiffness simply by replacing 
$e\rightarrow e^*$ and $l_B\rightarrow l^*$ in the expression for the 
electronic spin stiffness at $\nu=1$.\cite{sondhi} 
This corresponds to a CF mean-field approach,\cite{nakajima}
and $\mu(c)$ may thus be interpreted as a mismatch factor with respect to
the mean-field result.

Performing an analogous calculation for the quartic term of the Hamiltonian
(\ref{hintboso}), one finds
\begin{eqnarray}
\nonumber
\langle  \Hhat_{I} \rangle
  &=& {\cal N} \sum_{\mathbf{k,p,q}}
       v_0(k)\left[F_{0}^{CF}(k)\right]^2\sin(\mathbf{k}\wedge\mathbf{p}l^{*2}/2)
\\ \nonumber && \\ \nonumber
 &&    \times\sin(\mathbf{k}\wedge\mathbf{q}l^{*2}/2)
       \Omega^-_{\mathbf{k}+\mathbf{q}}\Omega^-_{\mathbf{p}-\mathbf{k}}
                         \Omega^+_{-\mathbf{p}}\Omega^+_{-\mathbf{q}} 
\\ \nonumber &&\\ \nn
 &\approx& \frac{(1-c^2)^2}{2}\int d^2r\,d^2r'\;V(\mathbf{r}-\mathbf{r'})
     q(\mathbf{r})q(\mathbf{r'}),\\
\label{energyskyrmion2}
\end{eqnarray}
where ${\cal N} = 2/\left(\mathcal{A}(4\pi)^4N^{*2}_{\phi}(l^*)^8\right)$. 
Therefore, using equations (\ref{energyskyrmion1}) and
(\ref{energyskyrmion2}), one obtains the average value of the total Hamiltonian
with respect to the state $|sk^{CF}\rangle$
\begin{eqnarray}
\nonumber
\langle \Hhat \rangle &=&
  \frac{1}{2}\rho^0_S\int d^2r\;\left[\nablab\mathbf{n}(\mathbf{r})\right]^2
+ \frac{1}{2}g^*\mu_BB\frac{1}{2\pi}N^*_\phi
\\ \nonumber && \\ \nn
 && - \frac{1}{2}g^*\mu_B\frac{1}{2\pi l^{*2}}\int d^2r\;\mathbf{n}\cdot\mathbf{B}
\\ \nonumber && \\ \label{energiask}
 && +\frac{e^2(1-c^2)^2}{2}\int d^2r\,d^2r'\;
     \frac{q(\mathbf{r})q(\mathbf{r'})}{|\mathbf{r}-\mathbf{r'}|}
\;\;\;\;
\label{Haverage}
\end{eqnarray}
Eq.\ (\ref{Haverage}) corresponds to an energy functional of a static 
configuration of the vector field $\mathbf{n}(\br)$, derived from a Lagrangian
density similar to (\ref{lagsondhi}). Notice that the coefficient of
the Zeeman term, 
$$
\frac{g^*\mu_B}{2} \frac{\nu^*}{ 2\pi l^{*2}} = 
\frac{g^*\mu_B}{2}\frac{\nu}{2\pi l_B^2},
$$ 
is identical to that of Sondhi {\it et al.},\cite{sondhi} 
with the difference that here, this term is
derived from first principles and not assumed phenomenologically. Furthermore,
the last term in Eq.\ (\ref{Haverage}) accounts for the Coulomb interaction between
CF-quasiparticles with effective charge $e^* = e(1-c^2)$.

The analysis of the semi-classical limit of the interacting bosonic
model (\ref{hintboso}) allows us to calculate analytically the  
spin stiffness. From Eq.\ (\ref{stiffness}), one obtains
\beq
\label{stiffness1}
\rho_S =   \left\{\begin{array}{r@{\quad}l} 
             6.207\times 10^{-3} & \nu=1/3, \\
             2.269\times 10^{-3} & \nu=1/5,
            \end{array} \right.
\eeq
in units of $e^2/\epsilon l_B$.

There is a set of different values for the spin stiffness at $\nu=1/3$
and $1/5$ available in the literature. 
Based on Ref.\ \onlinecite{mandal}, Mandal and Ravishankar\cite{mandal2}
estimated that the spin stiffnesses at $\nu=1/3$ and $1/5$ are equal to  
$1.591\times 10^{-3}$ and $4.774\times 10^{-4}\; e^2/\epsilon l_B$,
respectively. These values are derived from the 
spin stiffness at $\nu=1$ with the replacements $e\rightarrow e^*$ and
$l_B \rightarrow l^*$, and may be interpreted as the mean-field values of the 
CF spin stiffness, as indicated above. In fact, these results are in agreement 
with estimates by Sondhi {\it et al.}\cite{sondhi}
The spin stiffness has also been calculated numerically by
Moon {\sl et al.},\cite{moon} using the 
hypernetted-chain-approximation.\cite{macdonald2} In this case, it was
found that $\rho_S = 9.23\times 10^{-4}\;\;[\nu=1/3]$ and $2.34\times
10^{-4}\;e^2/(\epsilon l_B)\;\;[\nu=1/5]$. Those values are also considered
by Lejnell {\it et al.}\cite{lejnell} in the study of the quantum Hall 
skyrmion.
Furthermore, the spin stiffness at $\nu=1/3$ as a function of temperature 
and finite width has been estimated by Murthy\cite{murthy2} in the 
Hartree-Fock approximation of the Hamiltonian theory. At low
temperatures ($0 < T < 2\;K$), one finds $\rho_S(T) \approx 2.125\times
10^{-3}$. 

The values we obtain with the help of the bosonization approach in the 
Hamiltonian theory are larger than the theoretical estimates found in the
literature. In the case of the mean-field values obtained by Sondhi 
{\it et al.}\cite{sondhi} and Mandal and Ravishankar,\cite{mandal2} the 
mismatch is given by the factor $\mu(c)$, which is $3.9$ at $\nu=1/3$ and
$5.1$ at $\nu=1/5$. The origin of this mismatch is the Gaussian 
$\exp(-|ql_B|^2/2c^2)$ in the CF form factor (\ref{cfformfactor}), which
takes into account the difference between the magnetic length of the electron
and that of the pseudo-vortex, which constitute the CF.\cite{MS} 
The mismatch is thus
due to the internal structure of the CF, which is not taken into account in
a mean-field approximation, where the CF is considered as a pointlike 
particle with charge $e^*$. 

A comparison with experimental data has only been done in Ref.\
\onlinecite{murthy2}, where the temperature dependence of the spin 
polarization at $\nu=1/3$ has been evaluated, using the
continuum quantum ferromagnet model in the large-$N$ approach.
In this work, the two free parameters of the model (spin stiffness and 
magnetization density) have been taken temperature-dependent. This
formalism had previously been used to describe the spin polarization of
the 2DEG at $\nu = 1$,\cite{read} but with constant
parameters. The agreement between the theoretical result and the experimental
data,\cite{khandelwal2} even when only the zero-temperature values of two 
free parameters were taken into account, indicated that, within this
formalism, the estimated value of the spin stiffness is quite
reasonable. Although we find a larger value for the spin stiffness than
Murthy's, it can be reduced if finite-width effects are
included in the model. Remember that the preferred combination of the
CF density is less accurate at larger momenta and the inclusion of
finite-width effects softens the short-range terms of the interaction
potential.   

Finally, the energy of the CF skyrmion ($E_{sk}$) may also be 
estimated in the long-wavelength limit. The leading term of the skyrmion 
energy 
is $4\pi\rho_S$,\cite{sondhi} which is basically the energy of the skyrmion in 
the limit of zero Zeeman
energy. One obtains from the values (\ref{stiffness1})
\beq
\label{Esk1}
E_{sk} = \left\{\begin{array}{r@{\quad}l} 
                0.078 & \nu=1/3, \\
                0.028 & \nu=1/5,
               \end{array} \right.
\eeq
in units of $e^2/\epsilon l_B$.
These results are in agreement with 
the energy of a large (number of spin reversals) skyrmion
calculated by Kamilla {\it et al.}\cite{kamilla} This indicates that  
the hard-core wavefunction description for the CF skyrmion and the
Hamiltonian theory in the bosonization approach converge
in the long-wavelength limit.

\section{Summary}

In conclusion, we have calculated neutral spin excitations of FQHE systems at 
$\nu=1/(2s+1)$, using a bosonization approach for the Hamiltonian theory of 
the FQHE proposed by Murthy and Shankar.\cite{MS} The
generalization of this formalism with the inclusion of a discrete
degree of freedom has been presented, 
under the assumption that one may neglect higher CF levels, which are separated
from the ground state by a larger gap than the spin-level splitting 
($\gamma\rightarrow 0$ limit).

Because in the CF picture the electronic system at
$\nu=1/(2s+1)$ is described by a CF model with
effective filling factor $\nu^*=1$, we can apply the bosonization
method for the 2DEG at $\nu=1$, recently developed by two
of us.\cite{doretto} In this case, the neutral spin excitations of the
system are treated as bosons, and the original interacting CF model is
mapped onto an interacting bosonic one.

The dispersion relation of the bosons (neutral spin reversal
excitations) has been evaluated analytically for an ideal system, 
neglecting finite-width, Landau-level-mixing and disorder
effects. We have illustrated the results for FQHE systems at $\nu=1/3$ and
$\nu=1/5$. The bosonization approach allows us to calculate
analytically the energy of the spin excitations over the whole momentum
range. 

Our results agree with numerical investigations in  
the time-dependent Hartree-Fock approximation.\cite{murthy}
We have shown that the long-wavelength limit of the bosonic dispersion 
relation corresponds to spin-wave excitations
of the quantum Hall ferromagnet of CFs, which is the ground state
of the system. The energies of the spin waves are larger than those obtained in
previous theoretical studies by Nakajima and Aoki.\cite{nakajima}
The large-wavevector limit of the dispersion relation is related to the 
energy to create a CF-quasiparticle-quasihole pair ($E_{ph}$), the 
constituents of which are far apart from each other.
Our results are also larger than numerical ones obtained with the help of
trial wavefunctions for the excited state in the CF picture
\cite{mandal-jain} and exact-diagonalization studies on systems with a few
number of particles.\cite{rezayi2} 
Even if these values occur to be closer than ours to experimental 
estimates\cite{dujvone1,dujvone2} of the $E_{ph}$, an excellent agreement is
obtained when finite-width effects are taken into account. Indeed, the width
of the quantum well used in the experiments corresponds to a width parameter
$\lambda \sim 3l_B$ ($\sim 300\,$\AA ~at $7\,$T), for which we obtain a 
reduction
of $E_{ph}$ roughly by a factor of three. Note that finite-width results are 
more reliable in the Hamiltonian theory, which was originally derived in the 
long-wavelength limit, than that for the ideal 2D case because, in this case,
short-range contributions of the interaction potential are 
suppressed.\cite{murthy} 
This cuts off short-wavelength fluctuations and thus brings the model back 
to its regime of validity.

In the framework of the interacting bosonic model, we have shown that the
interaction potential between the bosons accounts for the formation of
two-boson bound states. As in the case of the 2DEG at $\nu=1$, 
the two-boson bound state is interpreted in terms of a small
CF-skyrmion-antiskyrmion pair
excitation. FQHE systems at $\nu=1/3$ and $1/5$ have been investigated. 
Based on the relation between the asymptotic limit of the energy
of the lowest two-boson bound state and $E_{ph}$, we conclude that
a CF quasiparticle dressed by spin waves is more stable than a bare
CF quasiparticle only at $\nu=1/3$. Experimental results at $\nu=1/5$,
such as Knight shift data, are lacking at the moment and may give further 
insight into the spin-excitation spectrum of this state.

Finally, we have shown that the semi-classical limit of the interacting
bosonic model yields an energy functional derived from a
Langrangian density, similar to the phenomenological one considered by
Sondhi {\it et al.}\cite{sondhi} for the study of the quantum Hall
skyrmion. We thus corroborate the relation between the two-boson bound state 
and small skyrmion-antiskyrmion pair excitations. 

\bigskip
\begin{acknowledgments}
RLD and AOC kindly acknowledge Funda\c{c}\~ao de Amparo \`a
Pesquisa de S\~ao Paulo (FAPESP) and Fundo de Apoio ao Ensino e \`a
Pesquisa (FAEP) for the financial support. AOC
also acknowledges support from Conselho Nacional de
Desenvolvimento Cient\'{\i}fico e Tecnol\'ogico (CNPq). 
One of us (PL) wishes to thank Henri Chambert-Loir for stimulating
discussions. 
MOG and CMS are supported by the Swiss National Foundation 
for Scientific Research under grant No. 620-62868.00.
\end{acknowledgments}


\begin{thebibliography}{99}
\bibitem{comment2} Although the bosonic representation of the CF spin
  density operators are not derived here, they are similiar to the
  ones derived in Ref.\ \onlinecite{doretto} for the 2DEG at $\nu=1$,
  without the Gaussian factors $\exp(-|ql_B|^2/4)$.
\end{thebibliography}
\end{document}